\documentstyle[11pt,psfig]{article}

\newcommand{\bee}{\begin{equation}}
\newcommand{\eee}{\end{equation}}
\newcommand{\ble}{\begin{equation}}
\newcommand{\ele}{\end{equation}}
\newcommand{\mper}{\mbox{\ .}}
\newcommand{\mcom}{\mbox{\ ,}}

\newcommand{\laur}{\marginpar{\raisebox{3in}{\small \tt LAUR 92-2249}}}

\title{
Can averaged orbits be used to extract scaling functions?
}
\author{
Ronnie Mainieri
}

\begin{document}

\maketitle
\laur

\begin{abstract}
Trajectory scaling functions are the basic element in the study
of chaotic dynamical systems, from which any long time average
can be computed.  It has never been extracted from an
experimental time series the reason being its sensitivity to
noise.   It is shown, by numerical simulations, that the
sensitivity of the scaling function is to drift in the control
parameters, and not noise.  It is also explained how naive
averaging of the orbit points may lead to erroneous results.
\end{abstract}

The experimental study of chaotic dynamical systems presents us with
complicated geometrical objects --- strange sets --- that have to be
simply characterized to be compared with theoretical predictions.  The
strange attractors are reconstructed from experimental time series
through a procedure known as phase space reconstruction
\cite{FarmerReconstruction,TakensReconstruction}, which determines the
strange attractor up to coordinate transformations.  Therefore any
characterization of the strange attractor must be independent of the
coordinates used.  Many different functions and sets of numbers have
been proposed to characterize strange attractors, such as $f(\alpha)$
spectrum of singularities \cite{FofAlpha} and fractal dimensions, but
the only complete characterization is the one given by the scaling
function \cite{FeigenbaumScaling}, defined later on.  There have been few
attempts to extract the scaling function from experimental data
\cite{FeigenbaumMashed,Experiment,ChristosScaling}, due mainly to its
sensitivity to noise in the system. In this paper I will make explicit
the difficulties and analyze a proposed extraction method: that of
averaging the behavior of the system in the reconstructed phase space
\cite{Experiment}.  I will show that the averaging procedure, as
proposed, is not an effective procedure to extract the scaling function
from an experimental time series.  The main difficulty is that although
averaging does reduce noise, it does not reduce the main source of
error in extracting the scaling function which is the detuning of the
external parameters from the ones where theory makes its predictions.

The scaling function $\sigma(t)$ was introduced by Feigenbaum
\cite{FeigenbaumScaling}, and gives the local contraction rate of an
asymptotically long periodic orbit after transversing a fraction $t$ of
the orbit.  From it all other quantities of physical relevance can be
explicitly computed.  Scaling functions should be contrasted to other
quantities that are extracted from dynamical systems, such as
generalized dimensions and $f(\alpha)$ spectrum of singularities.
Although these quantities are invariants of the dynamical system (they
remain unchanged if coordinates are changed), it is not possible to use
them to compute all physically observable quantities such as the
average energy dissipated in a chaotic circuit or correlations in the
time series.  The proof that the scaling function can be used to
compute all physical averages was given by Sullivan
\cite{SullivanMaximal} and also by Feigenbaum
\cite{FeigenbaumCMapPresentation}.

The approach to understanding the effects of noise and systematic
errors will be through the numerical simulation of circle maps.  I will
review the sine circle map in section \ref{CircleMaps} and give a few
of the definitions that will be used later on.  The various type of
errors that hinder the extraction of the scaling function from time
series are discussed in section \ref{Noise}.  Systematic errors will
also be discussed in that section, as they are the major source of
error in extracting the scaling function.  The details on how to
compute the scaling function are discussed in section
\ref{ScalingFunction}; in particular I will concentrate on how to
extract an approximation to the scaling function for the golden mean
rotation number.  All these sections are preliminaries to the results
discussed in section \ref{Simulations}, where the sine circle map with
noise in the parameters is explained.  The surprising result is that
very large noise levels have little effect on the scaling function when
compared to systematic errors.  In that section I also discuss a
non-ergodic behavior of circle maps that occurs while averaging.

\section{Circle maps}
\label{CircleMaps}

Maps of the circle occur whenever two oscillators are nonlinearly
coupled.  In general the asymptotic behavior of the coupled oscillators
can be well described by a map that gives the difference in phase
between them.  For a circle map there are two relevant parameters: one
which controls the ratio of the frequencies between the oscillators
when they are uncoupled ($\omega$ in equation (\ref{EqMap})), and the
other which controls the amount of coupling between the oscillators
($k$ in equation (\ref{EqMap})).  An example of a circle map that
arises from the study of Hamiltonian systems is the sine circle map:
\ble
	x_{i+1} = x_i + \omega - \frac{k}{2\pi} \sin(2\pi x_i)
	\mper
	\label{EqMap}
\ele
This is a map from the circle (parameterized from $0$ to $1$) to itself,
that is, all iterations of the map are computed $\mbox{}\bmod\, 1$.
This map models the phase difference between two coupled
oscillators.
The interesting property of coupled oscillators is that they can
mode-lock --- while one of the oscillators executes $p$ cycles, the
other goes through exactly $q$ cycles.  The fraction $p/q$ is the
rotation number of the map and it represents the average fraction
of the full range of the map 
transversed by each iteration.  In general the rotation number is
defined as
\bee
	\rho = \lim_{n\rightarrow \infty} \frac{x_n}{n}
	\mcom
\eee
with $x_n$ computed without the $\bmod\, 1$ after each iteration.
If the strength of the coupling is non-zero, then there is a connected
region in parameter space $(k,\omega)$ where the rotation number is
constant: the Arnold tongue.

As the strength of the coupling between the two oscillators
increases, larger ranges of $\omega$ are part of a tongue.  At $k=1$ almost
all values of $\omega$ belong to some tongue and the map is said to be
on the critical line.  If the rotation number $\rho$ of the map is an
irrational number then the orbit of the map will be chaotic due to an
(instant) period doubling cascade at the critical line.  This is only
proven for a class of irrational numbers with a particular
number-theoretic property: if we expand the rotation number into a
continued fraction expansion, then the terms of the expansion will not
grow faster than a given power.  If
\bee
	\rho =
	\frac{1}{
	   \displaystyle a_1 +
	   \frac{1}{
	   	\displaystyle a_2 + \cdots
	   }
        }
	=
	\langle a_1, a_2, \ldots \rangle
\eee
is an irrational number, 
then for constants $C$ and $\theta$, the coefficients $a_n$ of the
expansion are bounded 
\ble
	a_n < C \theta^n
	\mper
\eee
The simplest proof of this chaotic behavior is through a
renormalization group construction \cite{SinaiHerman} which is
simplest for the golden mean irrational $\rho_g= \langle 1,1,1, \ldots
\rangle$.  In what follows I will concentrate on the golden mean
rotation number.  At this rotation number, the behavior of the map can
be approximated by considering a sequence of maps with rotation number
given by the approximants $Q_n/Q_{n+1}$ of $\rho_g$ obtained by
truncating its continued fraction expansion.  One finds that $Q_0=1$,
$Q_1=2$, and $Q_{n+1}= Q_n + Q_{n-1}$ (the Fibonacci numbers).

\section{Noise}
\label{Noise}

Noise in a dynamical system can be present in many forms:  in
observations, in the state, or in the dynamics.  If the system evolves
deterministically under a map $F_r$ depending on a parameter $r$,
but the position (state) is not
measured accurately, then there is observational noise.  This correspond
to having the dynamics $x_{i+1} = F_r(x_i)$, but observing $x_i + \xi_i$
rather than $x_i$, where $\xi_i$ is a random variable (noise).  The
system may also evolve stochastically.   In this case the noise may
change the state at each time step
\bee
	x_{i+1} = F_{r}(x_i) + \xi_i
	\mcom
\eee
or it may change the dynamics
\bee
	x_{i+1} = F_{r+\xi_i}(x_i)
	\mper
\eee
Combinations of all three types may occur and in general all are
present in a laboratory experiment.

The scaling function is very sensitive to noise and to the parameter
values of the map, which has made it difficult to extract it from
experimental data or even from numerical simulations.  As the outcome
of most experiments with chaotic systems is a time series, I will
concentrate on how scaling functions are extracted from them.  The
simplest method to eliminate the error in the time series is by
averaging it over several periods.  Even though averaging
can diminish observational and state noise, it does not change the fact
that there are drifts in the experiment that lead to systematic
errors.  As we will see, averaging over periods does little to
diminish the error in computing the scaling function, as it does not
change the errors made in tuning the parameters to the golden mean.
Because observation noise and state noise can be made small in an
experimental setup (by care in the experiment or by period
averaging), I will only consider the effects of dynamical noise
and systematic errors.


In experiments with systems at the borderline of chaos, systematic
errors are the largest.  When collecting the data for the circle map the
experimentalist has to tune the parameters so that the rotation
number is exactly the golden mean.   The golden mean is an
irrational number and its associated tongue has no width, which
makes the tuning only approximate.   Then to extract the scaling
function, or even a simpler thermodynamic average such as the
$f(\alpha)$ spectrum of singularities, the longest
possible data set must be collected, which implies that the
parameters must be kept at the golden mean for a long time.
The time scale is determined with respect to the natural
frequency of the system, if it is a self-oscillator, or with
respect to the external frequency, if it is a forced oscillator.
In a typical experimental setup the parameters cannot be kept
tuned to golden mean rotation number and a slow drift in the
parameters can be detected.  This drift is interpreted in the
experiment as a systematic error.

If the rotation number is determined from the Fourier spectrum, then
its accuracy is low.  If $N$ data points are used in computing the
spectrum, then the rotation number, which is a frequency, is known to
an accuracy of order $1/N$.  Better techniques for computing the rotation
number have been developed which take into account that the orbit
points have a well defined ordering around the circle.  With these
techniques it is possible to determine the rotation number to an
accuracy of order $1/N^2$ \cite{EckeRayleigh}.

To convey an intuition on how sensitive an experiment can be to drift
consider the Rayleigh-B\'{e}rnard convection experiment performed in a
mixture of $^{3}$He and $^{4}$He at mili-Kelvin temperatures
\cite{EckeMaenoPRL,MaenoEcke}.  The data from this experiment was
collected for several days without interruption and the relevant
control parameter --- temperature --- was kept tuned to the golden mean
value to within 1 part in $10^5$.  Nevertheless 24 hour fluctuations on
the rotation number could be seen while the laboratory air-conditioner
was turned off.  Once the laboratory temperature was regulated other
fluctuations on the scale of an hour could be detected with the $1/N^2$
method.  Variation of the position in parameter space seems unavoidable
in any experimental setup.

\section{Scaling Function}
\label{ScalingFunction}

\begin{figure}
	\centerline{\psfig{file=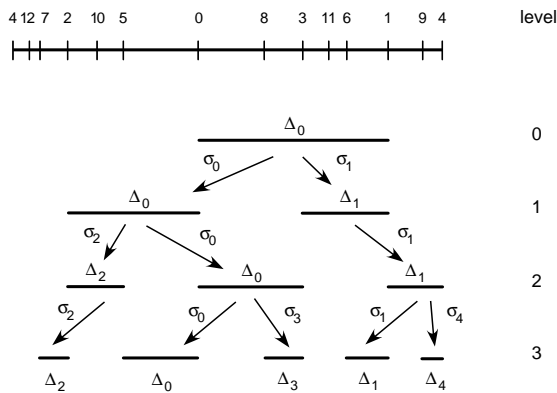,height=2.5in}}
	\caption[boo]{The segments used for the construction of the
	scaling function are determined from the first iterates
	of the map. The numbers $n$ on the top segment label the
	iterates $x_n$ of the map. By considering the orbit
	points separated by a Fibonacci number, the segments
	can be arranged in levels.}
	\label{FigSegments}
\end{figure}

The scaling function for a circle map is computed at a quadratic
irrational which has a periodic continued fraction expansion.  As the
inflection point, $x_0$, is iterated it rotates on average $\rho$, and
the points $x_0, x_1, \ldots$ of the orbit delimit a series of intervals or
segments along the circle.  The endpoints of these segments are not
two successive iteration points, such as $x_t$ and $x_{t+1}$, but
depend on how many times the initial point has been iterated
(see figure \ref{FigSegments}).   If the number of iterations is a
Fibonacci number, then the orbit points can be arranged in groups (or
levels) that recursively subdivide the circle into smaller and smaller
segments.  An example of the subdivision processes is given in figure
\ref{FigSegments}.   The first 13 orbit points of the golden mean
trajectory are indicated in the figure.  Notice that the segment
$\Delta^{(n)}_0$ is delimited by the orbit points $x_0$ and $x_{Q_n}$,
and that its location alternates to the right and left of the point
$x_0$.  The other segments of the level are determined by mapping the
segment $\Delta^{(n)}_0$ around the circle.  For universality, this
construction is not to be carried out with the actual map, but rather
with a $Q_n$ iterate of the map \cite{RandCircleMap,FeigenbaumCircleMap}.

For the case of a simple repeating number in the continued
fraction expansion, such as the golden mean, the segments are
given by
\bee
	\Delta_s^{(n)} = | x_s - x_{Q_n + s} |
\eee
from which we can define the values assumed by the scaling function at
different points
\bee
	\sigma^{(n)}_s =
	\frac{\displaystyle \Delta^{(n+1)}_s}
             {\displaystyle \Delta^{(n)}_s [s<Q_n] + 
			    \Delta^{(n)}_{s-Q_n} [s \geq Q_n]
	     }
	\mper
\eee
A square brackets \cite{KnuthConcrete} evaluates to one if the
expression within them is true and zero otherwise, so that the
denominator of the expression chooses one of the segments,
$\Delta^{(n)}_s$ or $\Delta^{(n)}_{s-Q_n}$, as appropriate for the
segment on the numerator (see figure \ref{FigSegments}).  An
approximation to the scaling function is obtained by the concatenation
of $Q_n$ short steps of length $1/Q_n$ and height $\sigma^{(n)}_s$ in
ascending order of $s$.  This defines a function from the unit interval
to itself.  The approximation in terms of steps of constant height is a
reasonable approximation because the variation in height of the steps
diminishes exponentially fast as the number of the level $n$
increases.  The construction of the continuous (and also
differentiable) almost everywhere scaling function is
\ble
	\sigma(t) = \lim_{n \rightarrow \infty}
		\sigma^{(n)}_{\lfloor t Q_n \rfloor}
	\mcom
	\label{EqLimit}
\ele
where $\lfloor x \rfloor$ is the function that gives the largest
integer smaller than $x$.
When evaluating the scaling function from a map with the
parameters different from the golden mean rotation number, then
there is another limit involved: that of approaching the
irrational number winding number.  The two limits do not commute, and
the irrational winding number must be approached before the limit
to an infinite number of levels.  In practice the irrational is
approached as well as possible.  Also, for universality, the
scaling function must be computed in a neighborhood of the
inflection point.  In practice the problem is
bypassed by taking $Q_0$ to be not 1, but a larger Fibonacci
number (see reference \cite{EckeRayleigh}).  This follows from the properties of circle maps with a
golden mean rotation number.  Under composition by a Fibonacci
number of times the orbit points accumulate around the starting
point for the iterations, which is taken to be the inflection point.

\begin{figure}
	\centerline{\psfig{file=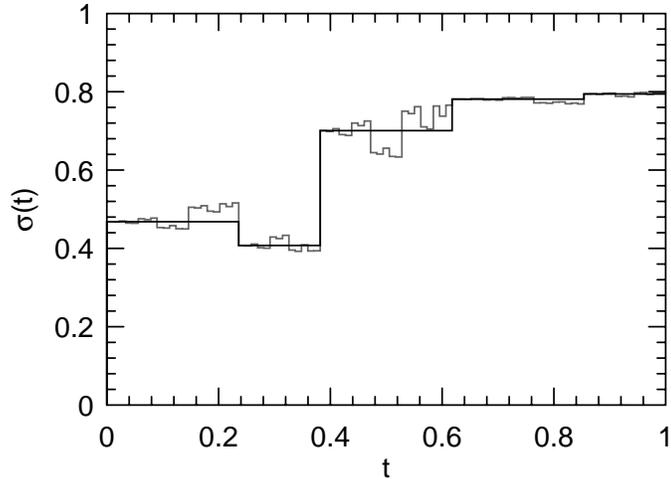,height=2.5in}}
	\caption[boo]{Three five step approximation (in black) to
	the limiting  scaling function (in gray).}
	\label{FigThree}
\end{figure}

The limit (\ref{EqLimit}) to compute the scaling function
has to be approximated in practice with a large enough $n$.
From experiments it has proven feasible to extract the 
scaling function which has 5 steps.  This approximation is
plotted in figure \ref{FigThree} together with
the limiting scaling function.

\section{Simulations}
\label{Simulations}

To understand the sources of error in determining the scaling function,
I will compute it from an orbit of a map that is not exactly at the
golden mean rotation number but at one of its continued fraction
approximants with a length typical of what is
obtained in laboratory experiments.  In particular I will consider the
orbit with rotation number $21/34$, which in the circle map occurs at
$\omega = 0.606\,439$ on the critical line ($k=1$).  To simulate the
effects of noise fluctuations the control parameters $k$ and $\omega$
of the sine circle map will be slightly varied at each iteration.  Both
$k$ and $\omega$ will be replaced by
\bee
   \begin{array}{lcr}
	k_i & =  & k + r_i \Delta k \\
	\omega_i & = & \omega + s_i \Delta \omega
   \end{array}
   \mcom
\eee
where $\Delta k$ and $\Delta \omega$ are the strength of the
fluctuations and $r_i$ and $s_i$ are random numbers uniformly
distributed in the interval from $-1$ to $1$.  Notice that $k$
and $\omega$ remain fixed during the random process, and for the
uniform distribution, represent the average values of $k_i$ and
$\omega_i$.  An orbit from the noisy sine circle map is generated by
\bee
	x_{i+1} = x_i + \omega_i - \frac{k_i}{2\pi}\sin ( 2\pi x_i)
	\mper
\eee
If the average parameter values are within the $21/34$ tongue, then
the map is iterated a few hundred times before any orbit points are used 
to compute the scaling function.  If the average parameters are not
within the tongue, then the map is started at the inflection point.

The orbit is averaged according to the procedure proposed by
Belmonte {\em et al.}\ \cite{Experiment}, where points that are nearby in
coordinate space are averaged together and coalesced into a
single orbit point of an averaged periodic orbit.  If the average parameter
values of the map are within the tongue of the rotation number
being considered, the group of points to be averaged can be unambiguously
distinguished for errors as large as $\Delta k = \Delta \omega =
0.05$, an error much larger than in most experiments.  If the
average parameter values are outside the tongue, then the number
of groups to be averaged will depend on the length of the data
set.

\begin{figure}
	\centerline{\psfig{file=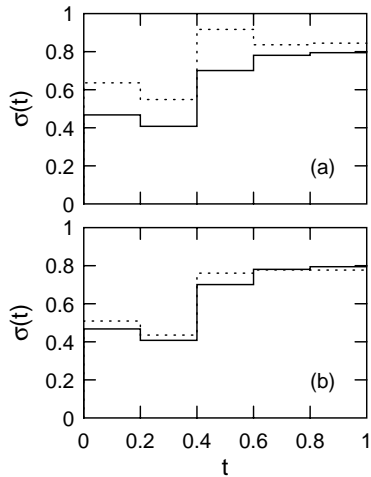,height=2.5in}}
	\caption[boo]{Different type of errors lead to different
	scaling functions.  In both figures the theoretical curve
	is indicated by a solid line.  In figure (a) the fluctuation errors
	are small, but the scaling function deviates largely from the
	theoretical curve.  In figure (b)
	the fluctuation errors are large, but the scaling function deviates
	only slightly from the theoretical curve.}
	\label{FigTwoscf}
\end{figure}

The first observation from the numerical simulations is that small
errors can lead to largely distorted scaling functions.  In figure
\ref{FigTwoscf}a the scaling function for a short orbit ($21/34$) at
parameters $k=0.9999$ and $\omega=0.6063$, which is close to the
superstable point of the tongue, is compared with the theoretical
curve.  The amplitude of the error fluctuations are small ($\Delta k =
\Delta \omega = 10^{-4}$) which keeps the map parameters within the
tongue.  In this case there are large deviation from the theoretical
curve.  In figure \ref{FigTwoscf}b, for the same orbit length, the
scaling function is computed with fluctuation noise 100 times larger,
but with parameters ($k=1.0$ and $\omega = 0.6066$) closer to the
golden mean critical point.  The difference between the scaling
function obtained from the short orbit and the theoretical curve is
smaller than in figure \ref{FigTwoscf}a.  This at first seems
paradoxical: the curve with larger fluctuations is closer to the
theoretical curve than the curve with smaller fluctuations.

To quantify the differences between the theoretical and short period
scaling function the $L^1$ norm can be used.  This norm is
proportional to the
area in between both curves.  If $\sigma(t)$ is the theoretical scaling
function with five steps, and $\sigma_o(t)$ is the scaling function
obtained from the orbit with $34$ points, also with five steps, then
the error between them is defined as
\bee
	e(\sigma,\sigma_o) = \frac{1}{c_o}\int_0^{1} dt\,
			| \sigma(t) - \sigma_o(t)|
	\mcom
\eee
where $c_0$ is a normalization constant.  The constant is chosen
so that error between the theoretical scaling function and the
one obtained from a short orbit at the irrational winding number
is one.  The constant $c_0$ is computed to be $0.01121$.
With this norm the error between the curves in figure
\ref{FigTwoscf}a is $2.26$ and between the the curves in figure
\ref{FigTwoscf}b is $0.53$.


%
\begin{figure}
	\centerline{\psfig{file=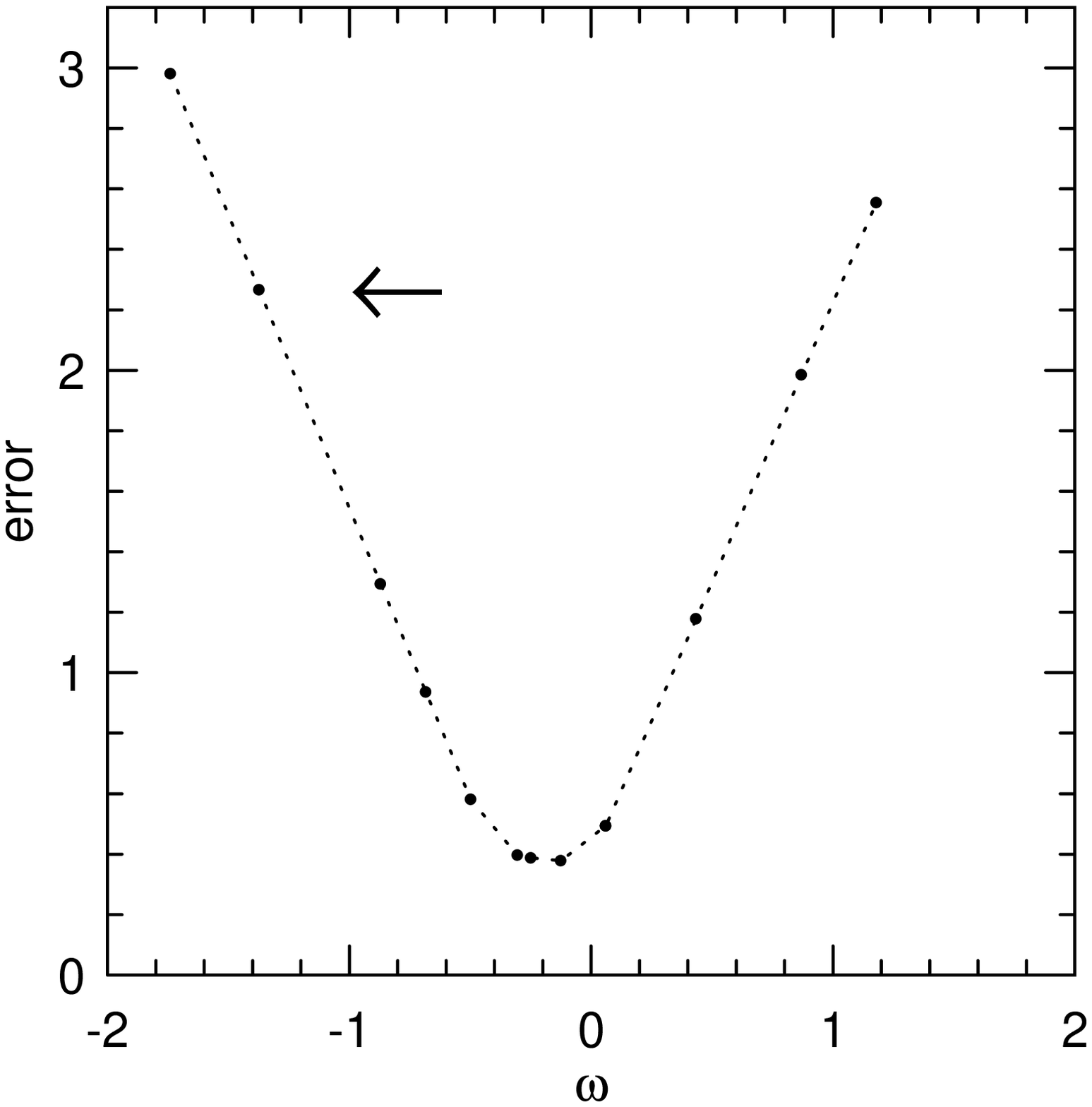,height=2.5in}}
	\caption[boo]{The error in approximating the scaling
	function by an orbit generated from a map with systematic
	error.  The winding number $\omega$ is measured in units
	of the width of the $21/34$ tongue away from the golden
	mean and the error is the area in between the curves.
	The arrow shows the location of the point for the $21/34$
	cycle}
	\label{FigError}
\end{figure}

The systematic error constitute the larger source of error.  This can
be verified by plotting the error between the scaling function
obtained at a point away from the golden mean rotation number and
another point along the critical line.  The further the rotation
number is from the golden mean, the larger the difference between the
two scaling functions.   In figure \ref{FigError} the inflection
point is iterated for 34 times; from this orbit a five step scaling
function is computed, which is then compared to the asymptotic five step
scaling function.  In the figure the rotation number is measured from
its departure from the golden mean rotation number in units of the
width of the $21/34$ tongue.  In actual units of the map the
horizontal axis ranges from $0.60638$ to $0.60685$, which is four times
the width of the $21/34$ tongue.  According to the plot, the error is
smallest when the rotation number is closest to the golden mean, and
increases as one departs from it on either side.   The exact minimum in
the error curve does not coincide with the golden mean because of finite size
effects in computing the scaling function.  For the error curve to be a
smooth function of the rotation number it is necessary that all orbits
start at the same point, the inflection point in this case.

The plot of figure \ref{FigError} was obtained from iterating a
map without fluctuations in the control parameters.  One would expect
the the results obtained without the error are those that would be
obtained had the map with error been iterated and averaged a large
number of times.  This is the case if the points are averaged
according to their time index, that is, for an orbit of period $P$, the
average over the fluctuations of the $k$-th point of a periodic orbit
are computed from
\bee
	\langle x_k \rangle
	=
	\lim_{n \rightarrow \infty } 
	   \frac{1}{n} \sum_{1 \leq i \leq n} x_{k + iP}
	\mper
\eee
But this may not be the average that is computed in an experiment.  Sometimes
it is simpler, or consistent with time delay coordinates, to average
the points that are close to each other in time delay space (this was
the procedure adopted in reference \cite{Experiment}).   In
table \ref{TblError} an orbit for a map at the superstable point has
been iterated with a small error ($\Delta k = \Delta \omega =
10^{-3}$).  The map is iterated while the
parameters fluctuate.   Each point is compared with the exact orbit and
iterated points that come close to the same exact orbit point are averaged
together.  From the averaged orbit the five level approximation to the
scaling function is computed and used to determine the error
associated with the orbit by comparing it to the scaling
function without noise.  By ``without noise'', I mean the scaling
function that is obtained by iterating the map with the
average parameters values of the simulation with noise.
The table shows the results of longer and longer averages.
At first the error diminishes, but as the number of samples
increases the error appears to remain constant.  
The conclusion
from the table is that one has to be careful that the limits involved
in the averaging procedure are well defined and converge to ones
expectations.
\begin{table}
  \begin{center}
	\begin{tabular}{|cl|}
	\hline \hline
	samples  &   \multicolumn{1}{c|}{error} \\ [0.5em]
	 $10^0$  &   $1.99065$ \\
	 $10^1$  &   $0.21849$ \\
	 $10^2$  &   $0.53230$ \\
	 $10^3$  &   $0.49462$ \\
	 $10^4$  &   $0.52383$ \\
	 $10^5$  &   $0.51801$ \\
	 \hline
	\end{tabular}
   \end{center}
   \caption[boo]{Error between the scaling function computed with
   and without noise.  The noisy map has fluctuating parameters with
   average at the superstable point of the $21/34$ tongue.  The
   averaging is done in coordinate space.  As the number of
   samples increases the error does not go to zero, as would be
   expected.}
   \label{TblError}
\end{table}

\section{Conclusions}

From the numerical simulations one sees that even large errors
can have little effect on the extraction of the universal scaling
function, {\em provided} that the parameters of the system are well tuned to the
golden mean rotation number at the transition to chaos, as can
be seen from figure \ref{FigTwoscf}.  The error in computing the
scaling function depends on how close the parameters are to the
transition point to chaos, a quantity that is difficult to
control in experiments, as they are invariably subject to drift.
The drift comes from the conflicting requirement of tuning the
parameters to the smallest possible tongue (and therefore at the
limit of instrumentation) and of obtaining the longest possible
times series.

Also from the numerical simulations the perils of averaging the
orbit in coordinate space where pointed out.  The noise in the
system causes the orbit to land close to the ``wrong'' group of
points for its phase within the period, which leads to the
non-convergence of the averaging procedure.  I have no
mathematical proof for this lack of convergence, but table
\ref{TblError} gives numerical evidence towards the result.

It hardly seems worthwhile to extract the scaling function given all
these difficulties, specially given that the $f(\alpha)$ spectrum of
singularities seems very robust to noise and simple to extract from
experimental time series. It also appears to give an infinity of scales
for the problem, just as the scaling function.  The difficulty with
this argument lies with the error bars of the spectrum of
singularities.  With error bars of the order of 1{\%}, the spectrum of
singularity is equivalent to just three of the values of the scaling
function \cite{FeigenbaumStrangeSets}; the spectrum of singularities
does not give individual scales but mixes them all into one function.
In order to extract further information from the spectrum of
singularities the errors would have to be reduced well below the 1\% level,
which does not seem possible even in numerical simulations.  Contrast
this with the scaling function.  There every different scaling (the
$\sigma^{(n)}_s$) can be individually and independently extracted.

Scaling functions are the fundamental objects in the study of low
dimensional dynamical systems, and it is surprising how little
attention they have received in the literature, both theoretically and
experimentally.  If they are to be extracted from experimental
time series new techniques will have to be developed to control
the systematic errors.

\section*{Acknowledgments}

I would like to thank 
Predrag Cvitanovi\'{c},
Robert Ecke,
Mitchell Feigenbaum,
and
Timothy Sullivan,
which over the years have clarified many aspects, theoretical and
experimental, in the study of chaotic systems in general and on scaling
functions in particular. I would also like to acknowledge
conversations with James Theiler.  This work was supported by a grant
from the Department of Energy.

\end{document}